\documentclass[aps,prb,reprint,superscriptaddress]{revtex4-2}
\usepackage{amsmath,amssymb}
\usepackage{graphicx}
\usepackage[colorlinks,linkcolor=blue,anchorcolor=blue,citecolor=blue,urlcolor=blue]{hyperref}
\usepackage{booktabs}
\usepackage{multirow}
\usepackage[textsize=tiny,backgroundcolor=yellow]{todonotes}

\begin{document}


\title{Anharmonicity and Coulomb pseudopotential effects on superconductivity in YH$_6$ and YH$_9$}



\author{Yucheng Ding}
\affiliation{International Center for Quantum Materials, Peking University, Beijing 100871, China}
\author{Haoran Chen}
\affiliation{International Center for Quantum Materials, Peking University, Beijing 100871, China}
\author{Junren Shi}
\email{junrenshi@pku.edu.cn}
\affiliation{International Center for Quantum Materials, Peking University, Beijing 100871, China}
\affiliation{Collaborative Innovation Center of Quantum Matter, Beijing 100871, China}


\date{\today}

\begin{abstract}
Anharmonic effects are widely believed to be the primary cause of the overestimation of superconducting transition temperatures of yttrium hydrides YH$_6$ and YH$_9$ in theoretical predictions. However, prior studies indicate that anharmonicity alone may be insufficient to account for this discrepancy. In this work, we employ the stochastic path-integral approach to investigate the quantum and anharmonic effects of ions in yttrium hydrides. Our calculations reveal significant corrections to the electron-phonon coupling parameters and an increase in the average phonon frequency compared to density functional perturbation theory, aligning closely with results from the stochastic self-consistent harmonic approximation. We find that properly taking into account the renormalization of the Coulomb pseudopotential due to the frequency cutoff, which is often overlooked in previous calculations, is critical to predicting transition temperatures consistent with experimental values for both YH$_6$ and YH$_9$. This indicates that, with this correction, anharmonic effects are sufficient to explain the discrepancies between experimental and theoretical results.
\end{abstract}


\maketitle

\section{Introduction}
Since the discovery of superconductivity in mercury, efforts have been focused on searching for superconductors with higher superconducting transition temperatures ($T_c$). The Bardeen-Cooper-Schrieffer (BCS) theory explains conventional superconductivity through electron-phonon coupling (EPC). It also illuminates the path towards higher $T_c$, which requires stronger EPC and higher phonon frequency. In this regard, hydrides are excellent candidates for achieving high $T_c$ due to the light mass of hydrogen atoms. Guided by this idea, various hydride superconductors have been theoretically predicted~\cite{hydride_review}. Over the past decade, multiple hydride superconductors, such as H$_3$S~\cite{H3S_the,H3S_exp}, LaH$_{10}$~\cite{LaH10_the,LaH10_exp1,LaH10_exp2}, CaH$_6$~\cite{CaH6_the,CaH6_exp1,CaH6_exp2}, YH$_{6}$ and YH$_{9}$~\cite{YH_the1,YH6_exp1,YH_exp2,YH_exp3}, have been successfully synthesized and exhibit $T_c$ exceeding 200 K, while Li$_2$MgH$_{16}$~\cite{LMH_the}, which is predicted to possess the highest $T_c$, has yet to be confirmed in experiments.

For hydride superconductors, discrepancies often arise between experimental and theoretical $T_c$ values. Several explanations have been proposed, including anharmonic effects on ion motion~\cite{H3S_SSCHA,CaH6_SSCHA,YH6_exp1}, diffusion of ion vacancies~\cite{LaH10_SPIA,CaH6_SPIA}, structure phase transitions~\cite{CaH6_phase,YH9_phase,LaH10_phase}, and nonhydrostatic stress in experiments~\cite{YH6_stress}. Among these mechanisms, the quantum and anharmonic effects of hydrogen ions are widely accepted as the cause for the overestimation of $T_c$ in theoretical calculations. One well-established method to incorporate these effects is the stochastic self-consistent harmonic approximation (SSCHA)~\cite{SSCHA_1,SSCHA_2,SSCHA_3}, which has been highly successful in predicting $T_c$'s for various hydride superconductors~\cite{H3S_SSCHA,CaH6_SSCHA}.

However, in the case of yttrium hydrides, the overestimation of $T_c$ in various calculations remains unexplained. YH$_6$ and YH$_9$ have been synthesized in multiple independent experiments, achieving maximum $T_c$ of approximately 220 K and 240 K, respectively. While theoretical calculations reproduce the dome-shaped $T_c$-pressure relationship observed in experiments~\cite{YH9_phase}, they overestimate $T_c$ by at least 20~K~\cite{YH6_exp1,YH_exp2,YH_exp3,YH_the1}. For instance, experimental $T_c$ values for \textit{Im$\bar{3}$m} YH$_6$ at 165 GPa and \textit{P$6_3$/mmc} YH$_9$ at 255 GPa are significantly lower than those predicted using density functional perturbation theory (DFPT). Even incorporating anharmonic effects via SSCHA fails to fully account for the discrepancy in YH$_6$~\cite{YH6_exp1}, and no systematic analysis of anharmonicity has been conducted for YH$_9$~\cite{YH9_anhar}.

In this paper, we present a first-principles study of anharmonic effects in yttrium hydride superconductors. To account for the anharmonicity and quantum fluctuations of hydrogen ions, we adopt the stochastic path-integral approach (SPIA) implemented with density functional theory (DFT)~\cite{SPIA,H_SPIA,H3S_SPIA}. SPIA is a non-perturbative approach that does not make assumptions about the nature of ion motion or the existence of an effective harmonic potential. Compared to SSCHA, SPIA offers broader applicability. It can be applied not only to solid systems~\cite{H3S_SPIA}, but also to systems with diffusive atoms, such as liquids~\cite{SPIA}, superionic solids~\cite{LMH_SPIA}, and systems with diffusive defects~\cite{LaH10_SPIA}. Using SPIA, we investigate the $T_c$ of YH$_6$ at 165 GPa and YH$_9$ at 255 GPa. We find that anharmonic effects significantly alter EPC parameters and increase the average phonon frequency, aligned with prior SSCHA findings. By properly taking into account the renormalization of the Coulomb pseudopotential due to frequency cutoff, which is neglected in earlier studies, we obtain $T_c$ values that closely match experimental observations for both compounds. This indicates that anharmonic effects are sufficient to explain the discrepancies between experimental and theoretical results in these systems.

The remainder of the paper is organized as follows. In Sec.~\ref{sec:SPIA}, we give a brief introduction of SPIA for anharmonic solids. In Sec.~\ref{sec:SC}, we apply SPIA to study the superconductivity of YH$_6$ and YH$_9$. The computational details are provided in Sec.~\ref{sec:computation}. We then present the results of our calculations and compare them with prior studies in Sec.~\ref{sec:result}. The renormalization of the Coulomb pseudopotential is discussed in Sec.~\ref{sec:mustar}. Finally, we summarize our results in Sec.~\ref{sec:summary}.

\section{Stochastic path-integral approach\label{sec:SPIA}}
SPIA is a method for determining $T_c$ of superconductors based on \textit{ab initio} path integral molecular dynamics (PIMD)~\cite{PIMD,AI_PIMD}. The key to this approach is to determine the effective interaction between electrons through the fluctuation of the electron-ion scattering $T$ matrix.

To study the electron-ion scattering, we start with the electron Green's function to describe the electron state. The PIMD simulation samples a set of ion configurations $\boldsymbol{R}(\tau)$ at each imaginary time $\tau$. For each sampling, the electron Green's function $\hat{\mathcal{G}}[\boldsymbol{R}(\tau)]$ can be determined with respect to the corresponding ionic field. The physical electron Green's function $\hat{\bar{\mathcal{G}}}$ is the average over each sampled instantaneous electron Green's function, i.e., $\hat{\bar{\mathcal{G}}}=\langle\hat{\mathcal{G}}[\boldsymbol{R}(\tau)]\rangle_C$.

The propagation and pairing of electrons occur in states that diagonalize the physical electron Green's function. For harmonic solids, this basis is just the Bloch states. For anharmonic solids, a generalized Bloch basis can be determined by diagonalizing the effective Hamiltonian which is derived from the electron Green's function in plane-wave basis~\cite{H3S_SPIA}. The index of the generalized Bloch states can be written as $1=(n,\boldsymbol{k},\omega_j)$, where $\boldsymbol{k}$ is the quasi-wave-vector, $n$ is the band index, and $\omega_j$ is the Fermionic Matsubara frequency.

We then determine the electron-ion scattering $T$ matrix and electron-electron interaction under the generalized Bloch basis. For each ion configuration in PIMD, the scattering $T$ matrix of electron is determined by the identity $\hat{\mathcal{T}}[\boldsymbol{R}(\tau)]=\hbar\hat{\bar{\mathcal{G}}}^{-1}(\hat{\mathcal{G}}[\boldsymbol{R}(\tau)]-\hat{\bar{\mathcal{G}}})\hat{\bar{\mathcal{G}}}^{-1}$. The pair scattering amplitude $\hat{\Gamma}$ is the fluctuation of the electron-ion scattering $T$ matrix, given by $\Gamma_{11^\prime}=-\beta\langle\left|\mathcal{T}_{11^\prime}[\boldsymbol{R}(\tau)]\right|^2\rangle_C$. Finally, the effective interaction $\hat{W}$ can be determined by solving the Bethe-Salpeter equation
\begin{equation}
	W_{11^\prime}=\Gamma_{11^\prime}+\frac{1}{\hbar^2\beta}\sum_2W_{12}\left|\bar{\mathcal{G}}_2\right|^2\Gamma_{21^\prime}.
	\label{eq:W}
\end{equation}

For conventional superconductors, it is usually reasonable to apply the isotropic approximation~\cite{YH_the2}. Under this approximation, we replace the effective interaction $\hat{W}$ with an isotropic average over the Fermi surface, which is known as the EPC parameter $\lambda$ given by
\begin{equation}
	\lambda(j-j^\prime)=-N(\epsilon_F)\langle W_{n\boldsymbol{k},n^\prime\boldsymbol{k}^\prime}(j-j^\prime)\rangle_{\mathrm{FS}},
	\label{eq:lambda}
\end{equation}
where $N(\epsilon_F)$ is the electron density of states (DOS) at the Fermi level, and $\langle\cdots\rangle_{\mathrm{FS}}$ denotes the average for all initial states $(n\boldsymbol{k})$ and final states $(n^\prime\boldsymbol{k}^\prime)$ over the Fermi surface. The EPC parameters then enter the isotropic linearized Eliashberg equation~\cite{Allen-Dynes} which we use to determine $T_c$ in this paper
\begin{equation}
	\rho\Delta_j=\sum_{j^\prime}\left[\lambda(j-j^\prime)-\mu^*-\frac{\hbar\beta}{\pi}|\tilde{\omega}(j)|\delta_{jj^\prime}\right]\Delta_{j^\prime},
	\label{eq:Eliashberg}
\end{equation}
where $\tilde{\omega}(j)$ is the renormalized Matsubara frequency under isotropic approximation given by $\tilde{\omega}(j)=\frac{\pi}{\hbar\beta}(2j+1+\lambda(0)+2\sum_{i=1}^j\lambda(i))$. To include the effect of the Coulomb interaction between electrons, we introduce the Morel-Anderson pseudopotential $\mu^*$~\cite{mustar,mustar_GW}. Eq.~\eqref{eq:Eliashberg} is an eigenvalue problem to be solved for the eigenvalue $\rho$ and renormalized gap $\Delta_j$. The emergence of a positive $\rho$ indicates the instability towards forming Cooper pairs and thus the superconducting phase. It is shown in Ref.~\cite{SPIA} that the linearized Eliashberg equation remains valid for general systems including anharmonic solids as long as the effective electron-electron interaction is determined by Eq.~\eqref{eq:W} when calculating the EPC parameters.

When evaluating the EPC parameters in Eq.~\eqref{eq:lambda}, $N(\epsilon_F)$ is determined by averaging the DOS over all PIMD configurations on a denser $\boldsymbol{k}$-grid to obtain converged values (see Appendix~\ref{app:k-grid}). The average of $\hat{W}$ is weighted by the Lorentzian function $\gamma/[(\epsilon_{n\boldsymbol{k}}-\epsilon_F)^2+\gamma^2]$, where $\epsilon_{n\boldsymbol{k}}$ is the eigen-energy of the generalized Bloch states. The half-width of the Lorentzian is set to $\gamma=\tilde{\omega}(0)=\frac{\pi}{\hbar\beta}(1+\lambda(0))$~\cite{SPIA}.

When solving the linearized Eliashberg equation Eq.~\eqref{eq:Eliashberg}, we need to set a truncation $N$ for the summation over $j$. This truncation $N$ (i.e., the Matsubara frequency cutoff) results in a renormalization of $\mu^*$, as pointed out by Allen and Dynes~\cite{Allen-Dynes}. It is given by
\begin{equation}
	\frac{1}{\mu^*(N)}=\frac{1}{\mu^*}+\ln\frac{\nu_{\mathrm{ph}}}{\omega_N},
	\label{eq:mustar}
\end{equation}
where $\nu_{\mathrm{ph}}$ is the phonon cutoff frequency at which $\mu^*$ is evaluated~\cite{McMillan,mustar_GW}. This renormalized $\mu^*(N)$, which replaces $\mu^*$ in Eq.~\eqref{eq:Eliashberg}, is crucial for obtaining a $T_c$ that is independent of the truncation $N$ (see Sec.~\ref{sec:mustar}).

\section{Superconductivity of YH$_6$ and YH$_9$\label{sec:SC}}
\subsection{Computational details\label{sec:computation}}
\begin{figure}[htbp]
	\centering
	\includegraphics[width=0.49\textwidth]{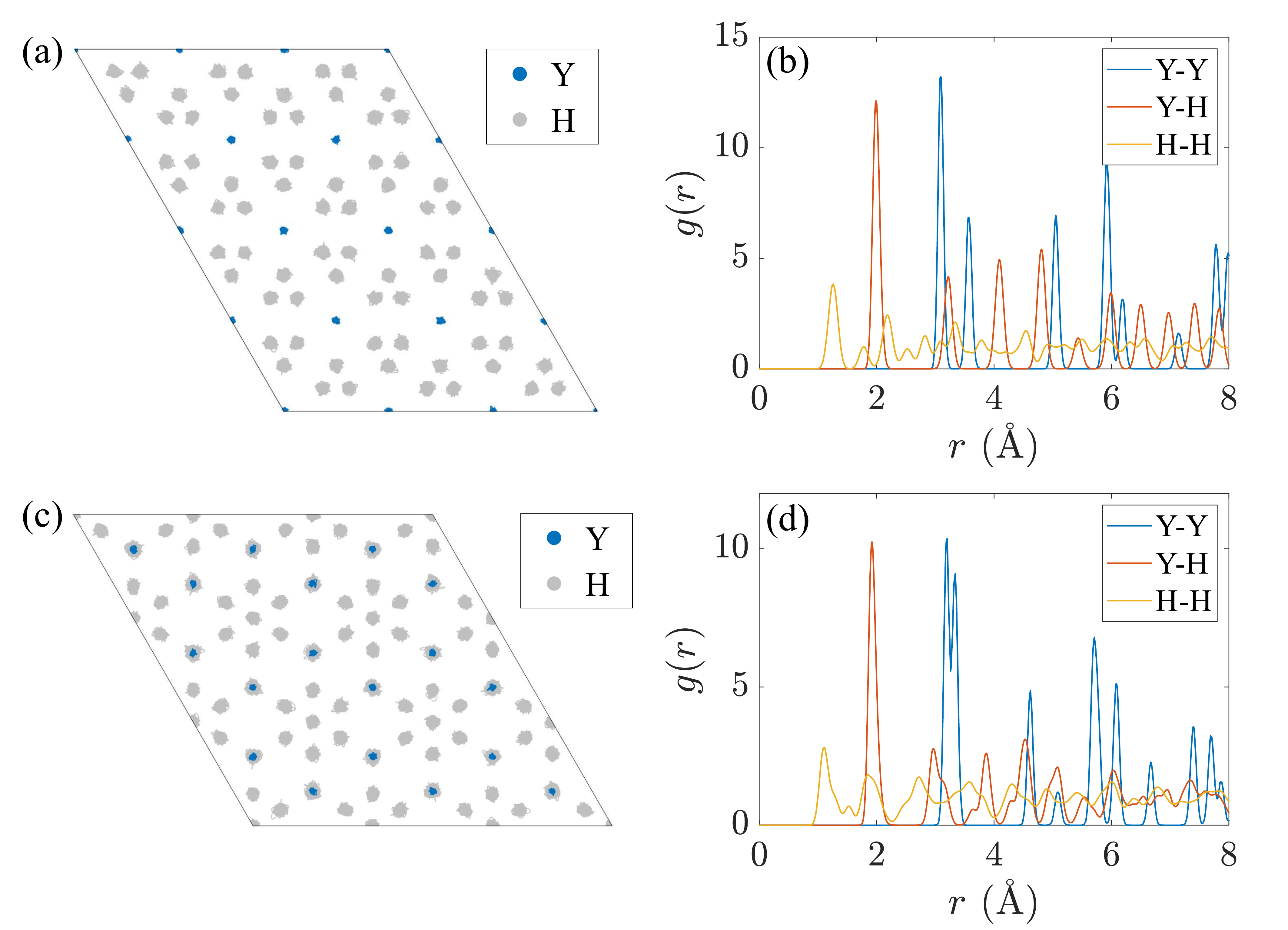}
	\caption{Trajectories of centroid mode of all hydrogen (gray) and yttrium (blue) ions in PIMD simulation for (a) YH$_6$ under 165 GPa at 220 K from [100] view, and (c) YH$_9$ under 255 GPa at 240 K from [001] view. (b) and (d) are corresponding RDFs.}
	\label{fig:PIMD}
\end{figure}

The crystal structures of YH$_6$ of \textit{Im$\bar{3}$m} space group at 165 GPa and YH$_9$ of \textit{P$6_3$/mmc} space group at 255 GPa are taken from Refs.~\cite{YH6_exp1,YH_exp2,YH_the1} and structurally optimized. All PIMD and DFT calculations are performed using a modified version of the Vienna \textit{ab initio} Simulation Package (VASP) code~\cite{VASP}. The projector-augmented wave (PAW) method~\cite{PAW} is used to describe the ion-electron interaction, and the Perdew-Burke-Ernzerhof (PBE) functional~\cite{PBE} is used to describe the exchange-correlation effect of electrons.

To accelerate the PIMD simulation while maintaining its first-principles accuracy, we apply the machine learning force field (MLFF) module of VASP~\cite{MLFF_1,MLFF_2,MLFF_3}. The modified program allows an on-the-fly training of MLFF in (PI)MD simulations~\cite{LMH_SPIA}. The details of training can be found in Appendix~\ref{app:MLFF_train}.

Using the MLFF generated above, we perform PIMD simulations for YH$_6$ and YH$_9$ with bead number $N_b=16$. A Langevin thermostat with the friction coefficients of centroid mode set to [10 10] ps$^{-1}$ is used to control the temperature of the canonical (NVT) ensemble~\cite{Langevin}. We use a $3\times3\times4$ supercell containing 252 atoms for YH$_6$ and a $3\times3\times2$ supercell containing 360 atoms for YH$_9$. YH$_6$ is simulated at 220 K while YH$_9$ is simulated at 240 K. The overall simulation time is 3 ps with a time step of 0.5 fs. 

The ion configurations of YH$_6$ and YH$_9$ are uniformly sampled with a spacing of 40 time steps after a 0.5-ps equilibrium. DFT calculations are performed for these configurations. An energy cutoff of 350 eV for plane waves is used to expand electron wave functions, and a $3\times3\times3$ $\Gamma$-centered $\boldsymbol{k}$-point grid is used to sample the Brillouin zone of the supercell. The converged DFT results are then used as inputs of our MATLAB implementation of SPIA~\cite{H3S_SPIA} to reconstruct the Hamiltonian. Finally, the effective electron-electron interaction $\hat{W}$ is calculated on the irreducible wedge of a $6\times6\times6$ $\boldsymbol{k}$-point grid of the supercell. The EPC parameters $\lambda(m)$ are calculated by averaging $\hat{W}$ near the Fermi surface. The half-width of the Lorentzian smearing is 0.16 eV for YH$_6$ and 0.18 eV for YH$_9$.

\subsection{SPIA results\label{sec:result}}
The results of PIMD simulations are shown in Fig.~\ref{fig:PIMD}. Figures~\ref{fig:PIMD}(a) and \ref{fig:PIMD}(c) illustrate the trajectories of hydrogen and yttrium ions in YH$_6$ and YH$_9$, respectively. Their radial distribution functions (RDFs) are shown in Figs.~\ref{fig:PIMD}(b) and \ref{fig:PIMD}(d). It can be seen that both hydrogen and yttrium ions vibrate near their equilibrium positions without diffusion, and their RDFs exhibit sharp peaks. This indicates that both YH$_6$ and YH$_9$ are solid at the simulated temperature. On the other hand, hydrogen ions exhibit relatively large vibration amplitudes and a broadened H-H distribution, which suggests potential strong anharmonicity.

\begin{figure}[htbp]
	\centering
	\includegraphics[width=0.5\textwidth]{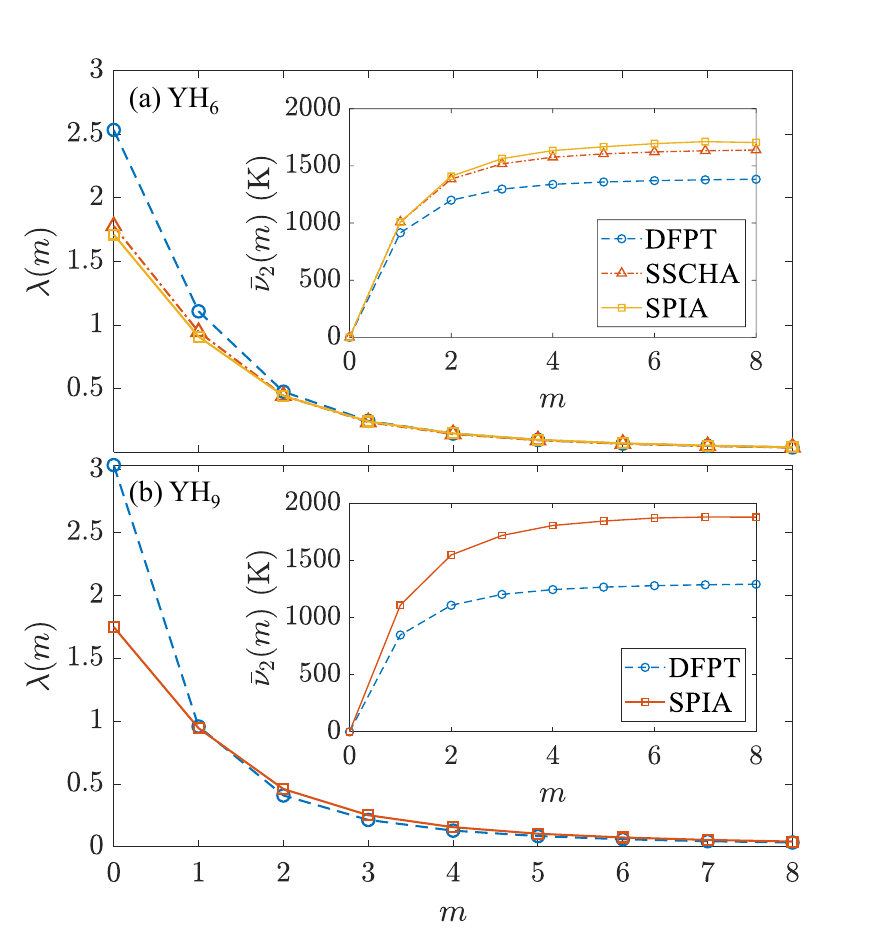}
	\caption{(a) EPC parameters $\lambda(m)$ of YH$_6$ under 165 GPa at 220 K. DFPT and SSCHA results are calculated using the Eliashberg spectral functions presented in Ref.~\cite{YH6_exp1}. (b) EPC parameters $\lambda(m)$ of YH$_9$ under 255 GPa at 240 K. DFPT result is calculated using the Eliashberg spectral function presented in Ref.~\cite{YH9_phase} under 260 GPa. The inset shows the asymptotic behavior of $\bar{\nu}_2(m)=2\pi/\hbar\beta\sqrt{m^2\lambda(m)/\lambda(0)}$, whose asymptotic value gives the average phonon frequency $\bar{\nu}_2$.}
	\label{fig:lambda}
\end{figure}

The EPC parameters of YH$_6$ and YH$_9$ are shown in Fig.~\ref{fig:lambda}. For YH$_6$, the SPIA results exhibit a significant suppression of $\lambda(0)$, $\lambda(1)$ and $\lambda(2)$ compared to the DFPT results. The values of these parameters, the behavior of $\bar{\nu}_2(m)=2\pi/\hbar\beta\sqrt{m^2\lambda(m)/\lambda(0)}$ and its asymptotic value (i.e., the average phonon frequency $\bar{\nu}_2$) are close to the results of SSCHA, which is shown in Fig.~\ref{fig:lambda}(a) and Table~\ref{tab:Tc}.

\begin{table}[tbp]
	\centering
	\begin{ruledtabular}
		\begin{tabular}{ccccccc}
			System & Method & $\lambda(0)$ & $\lambda(1)$ & $\lambda(2)$ & $\bar{\nu}_2$ (K) & $T_c$ (K)\\
			\midrule
			\multirow{3}{*}{YH$_6$} & DFPT & 2.534 & 1.109 & 0.477 & 1399 & 270 (248) \\
			& SSCHA & 1.779 & 0.948 & 0.446 & 1662 & 245 (218) \\
			& SPIA & 1.712 & 0.910 & 0.444 & 1738 & 212 \\
			\midrule
			\multirow{2}{*}{YH$_9$} & DFPT & 3.038 & 0.958 & 0.409 & 1308 & 252 (236) \\
			& SPIA & 1.749 & 0.944 & 0.460 & 1905 & 240 \\
		\end{tabular}
	\end{ruledtabular}
	\caption{The first few EPC parameters $\lambda(m)$, average phonon frequency $\bar{\nu}_2$ and predicted $T_c$ of YH$_6$ and YH$_9$. We set $\mu^*=0.11$ for all methods. For DFPT and SSCHA results, their EPC parameters and average phonon frequencies are calculated based on the Eliashberg spectral functions presented in Refs.~\cite{YH6_exp1,YH9_phase}. The $T_c$ values are also directly taken from the same sources. $T_c$ values in parentheses are calculated using our method with renormalized $\mu^*(N)$. The DFPT result for YH$_9$ is for 260 GPa.}
	\label{tab:Tc}
\end{table}

For YH$_9$, the SPIA results exhibit a strong suppression of $\lambda(0)$, a slight suppression of $\lambda(1)$ and an enhancement of $\lambda(m)\ (m\geq2)$ compared to the DFPT results. This leads to a notable enhancement of $\bar{\nu}_2(m)$ values and thus the average phonon frequency $\bar{\nu}_2$, as shown in Fig.~\ref{fig:lambda}(b) and Table~\ref{tab:Tc}. While the suppressed $\lambda(1)$ would lower $T_c$, its effect is compensated by the enhancement of $\lambda(m)\ (m\geq2)$. This could explain why DFPT still yields a decent prediction of $T_c$ despite the sizable anharmonic corrections.

\begin{figure}[htbp]
	\centering
	\includegraphics[width=0.5\textwidth]{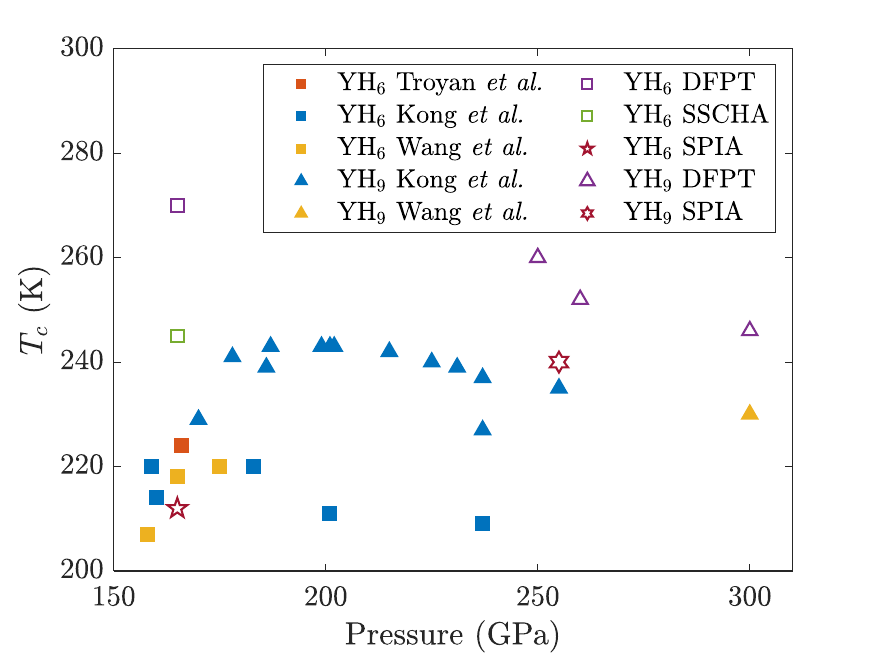}
	\caption{$T_c$ values of YH$_6$ and YH$_9$ at various pressures. Filled points indicate experimental data from Refs.~\cite{YH6_exp1,YH_exp2,YH_exp3}. Empty points represent theoretical predictions from Refs.~\cite{YH6_exp1,YH9_phase} and our SPIA calculations. $\mu^* = 0.11$ is used in all calculations.}
	\label{fig:Tc}
\end{figure}

Setting $\mu^* = 0.11$, as determined from $GW$ methods~\cite{YH_the2,mustar_GW}, and $\nu_{\mathrm{ph}}=2518$~K~\cite{mustar_GW}, we find $T_c$ to be 212~K for YH$_6$ and 240~K for YH$_9$. Table~\ref{tab:Tc} lists the results obtained using different methods. Figure~\ref{fig:Tc} summarizes observed or predicted $T_c$ values from various experiments and calculations. It can be seen that our SPIA predictions agree well with the experimentally observed $T_c$ values.

\subsection{Renormalization of the Coulomb pseudopotential\label{sec:mustar}}
\begin{figure}[htbp]
	\centering
	\includegraphics[width=0.49\textwidth]{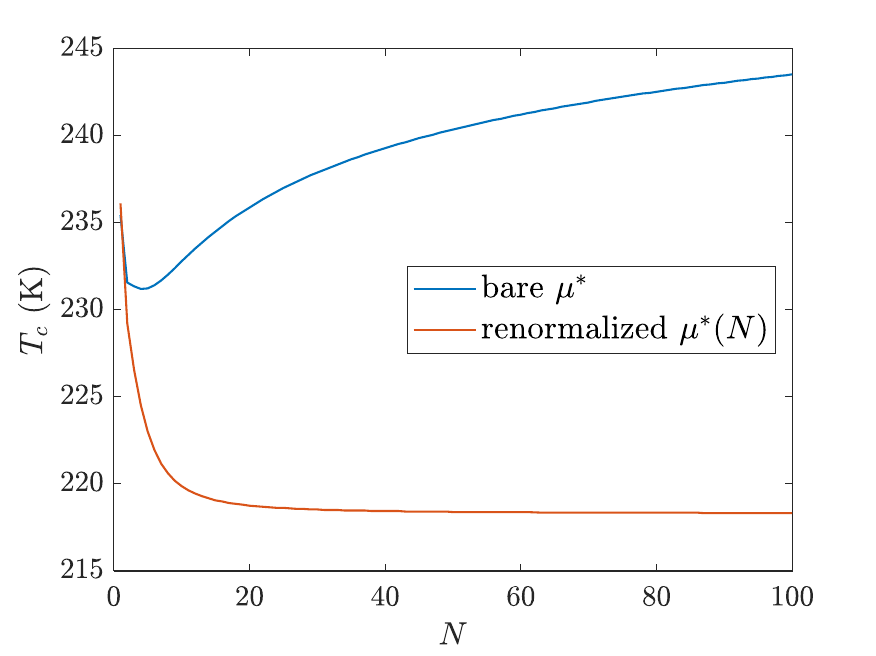}
	\caption{The relation between $T_c$ and frequency cutoff when solving the linearized Eliashberg equation. The truncation $N$ denotes the frequency cutoff at $\omega_N=(2N+1)\pi/\hbar\beta$. The blue line is calculated with the bare $\mu^*=0.11$ while the red line is calculated with the renormalized $\mu^*(N)$ defined in Eq.~\eqref{eq:mustar}. The Eliashberg spectral function used in calculation is the SSCHA result of YH$_6$ from Ref.~\cite{YH6_exp1}.}
	\label{fig:mustar}
\end{figure}

Properly taking into account the renormalization of $\mu^*$ due to the frequency cutoff in the Eliashberg equation, as specified in Eq.~\eqref{eq:mustar}, is crucial for correctly evaluating $T_c$~\cite{Allen-Dynes}.
In Fig.~\ref{fig:mustar}, we test the relation between $T_c$ and frequency cutoff by solving the linearized Eliashberg equation with and without the $\mu^*$ renormalization using EPC parameters of YH$_6$ calculated by SSCHA in Ref.~\cite{YH6_exp1}. It can be seen that without the $\mu^*$ renormalization, $T_c$ increases continuously with the frequency cutoff and does not show a trend of convergence even at large $N$ values. In contrast, with the $\mu^*$ renormalization, $T_c$ converges quickly. This indicates that only with the renormalization can we obtain a $T_c$ that is independent of the truncation.

The renormalization of $\mu^*$ plays an important role in our accurate predictions of $T_c$'s. As shown in Table~\ref{tab:Tc}, SPIA yields EPC parameters that are close to the SSCHA results for YH$_6$, yet it leads to a significant suppression of $T_c$. This can be attributed to the renormalization, which increases $\mu^*$. Table~\ref{tab:Tc} also includes $T_c$'s calculated using the renormalized $\mu^*$ for both DFPT and SSCHA methods (values in parentheses). These results reveal that, had the renormalization of $\mu^*$ been considered, SSCHA would have made a satisfactory prediction for the $T_c$ of YH$_{6}$. This suggests that anharmonic effects, combined with $\mu^*$ renormalization, are sufficient to bridge the gap between theory and experimental observations.

\section{Summary and discussions \label{sec:summary}}
In conclusion, we apply first-principles SPIA to study the anharmonic effects in yttrium hydride superconductors YH$_6$ and YH$_9$. The anharmonic effects suppress the EPC parameters $\lambda(0)$ and $\lambda(1)$ while increasing the average phonon frequency, which is in close agreement with previous SSCHA results. Combined with the renormalization of $\mu^*$ due to the frequency cutoff, which is neglected in previous calculations, it leads to $T_c$ that are consistent with the experimental ones.

Moreover, it would be interesting to investigate the dome-shaped $T_c$-pressure relationship of YH$_6$ and YH$_9$ with SPIA in future studies. Prior studies have attributed this non-monotonic behavior to structural phase transitions~\cite{YH9_phase} or changes in hydrogen vacancy concentration~\cite{LaH10_SPIA}. SPIA is well-suited for systems with structural distortions or hydrogen vacancy diffusion.

Given the critical role of the Coulomb interaction in predicting $T_c$, it would be desirable to determine its effect directly within the SPIA framework and without relying on the $\mu^*$ approximation. This could be achieved by generalizing state-of-the-art first-principles methods, such as DFT for superconductors (SCDFT)~\cite{SC_method} and $GW$-based methods~\cite{YH_the2,mustar_GW}, to a broader class of systems suitable for SPIA. Intriguingly, the ensemble average in SPIA may introduce additional corrections arising from the interplay between interactions and dynamic structural disorder. This will be left for future studies.

\begin{acknowledgments}
We acknowledge Xuesong Hu and Hao Jin for useful discussions. This work is supported by the National Key R\&D Program of China under Grant No. 2021YFA1401900 and the National Science Foundation of China under Grant No. 12174005.
\end{acknowledgments}

\appendix
\section{Convergence of calculation\label{app:k-grid}}
In Eq.~\eqref{eq:lambda}, the DOS $N(\epsilon_F)$ and effective electron-electron interaction $\hat{W}$ are calculated on a $\boldsymbol{k}$-point grid to sample the Brillouin zone of the primitive cell. To obtain converged EPC parameters, we need to perform non-self-consistent calculations of $N(\epsilon_F)$ and $\hat{W}$ on a denser $\boldsymbol{k}$-grid than that used for the electronic structure calculations.

\begin{figure}[htbp]
	\centering
	\includegraphics[width=0.5\textwidth]{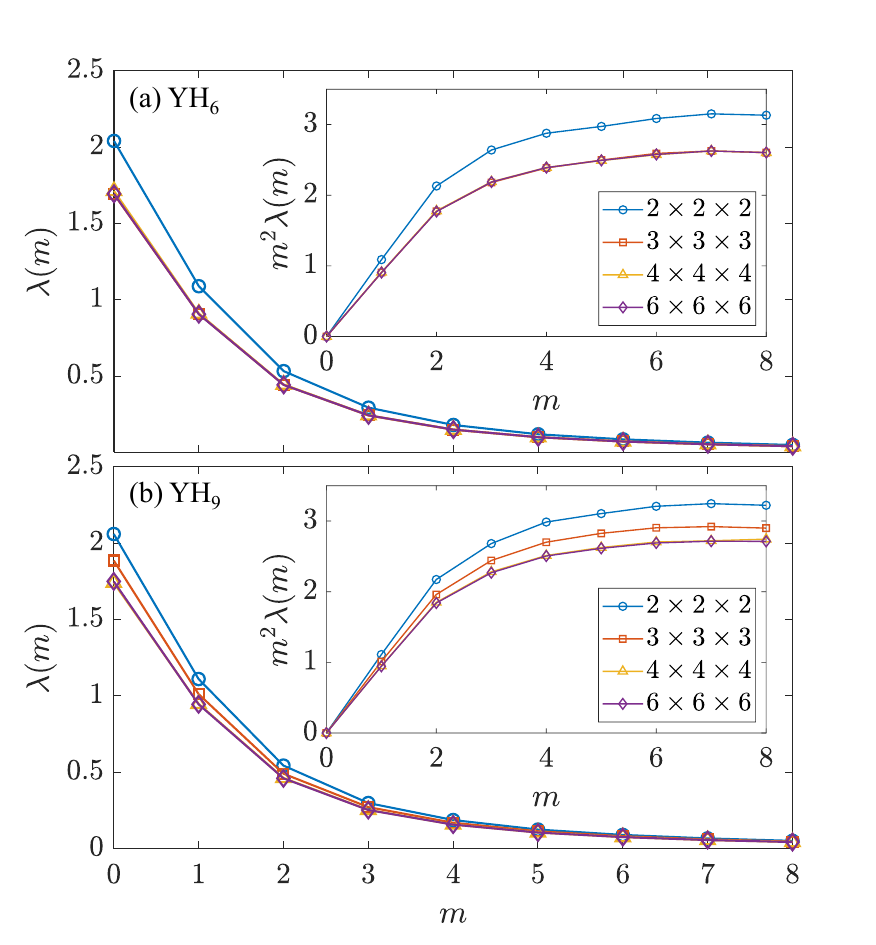}
	\caption{EPC parameters $\lambda(m)$ of (a) YH$_6$ and (b) YH$_9$ with respect to the $\boldsymbol{k}$-grid density of the supercell. The asymptotic behavior of $m^2\lambda(m)$ is shown in the inset.}
	\label{fig:k-grid}
\end{figure}

We test the $\boldsymbol{k}$-grid density of the supercell required for converged EPC parameters in Fig.~\ref{fig:k-grid}. It can be seen that a $4\times4\times4$ $\boldsymbol{k}$-grid yields EPC parameters and asymptotic values nearly identical to those of the $6\times6\times6$ $\boldsymbol{k}$-grid. Thus, a $6\times6\times6$ $\boldsymbol{k}$-grid of the supercell is sufficient for the convergence of calculations.

\section{Training of machine learning force field\label{app:MLFF_train}}
The training parameters are as follows. We use a $3\times3\times3$ supercell containing 189 atoms for YH$_6$ and a $2\times2\times2$ supercell containing 160 atoms for YH$_9$. The (PI)MD simulations are performed in the NVT ensemble with a Langevin thermostat to control the temperature. The friction coefficients of centroid mode are set to [10 10] ps$^{-1}$ and a time step of 1 fs is used. In DFT calculations, we use an energy cutoff of 500 eV for planes waves to expand electron wave functions and a $3\times3\times3$ $\Gamma$-centered $\boldsymbol{k}$-point grid to sample the Brillouin zone of the supercell. The radial and angular descriptors of MLFF are constructed using parameters ML\_SION=0.3, ML\_LMAX2=4, ML\_MRB=12, ML\_RCUT1=8 and ML\_RCUT2=5.

The training is performed in two steps. First, an on-the-fly MD training is performed. The temperature is raised from 200 K to 300 K in a 10-ps simulation. Second, quantum effects are taken into account by performing a PIMD simulation with bead number $N_b=8$. A 5-ps simulation is performed with the temperature raised from 250 K to 300 K. During the simulation, if the error of a structure exceeds a certain threshold, a DFT calculation is performed and the data is added to the training set.

The training results are as follows. For YH$_6$, 407 structures and 278 (1983) local configurations for Y (H) are chosen, and the root-mean-square error (RMSE) of the force reaches 0.051 eV/\AA. For YH$_9$, 349 structures and 284 (2438) local configurations for Y (H) are chosen, and the RMSE of the force reaches 0.071 eV/\AA.

\nocite{*}
\bibliography{reference}

\end{document}